\documentclass[fleqn,twoside]{article}
\usepackage[headings]{espcrc2}

\readRCS
$Id: espcrc2.tex,v 1.2 2004/02/24 11:22:11 spepping Exp $
\usepackage{graphicx}
\usepackage[figuresright]{rotating}
\usepackage{epsfig}
\title{Status of International Lattice Data Grid -- An Overview --}

\author{Akira Ukawa\address{Center for Computational Sciences, 
University of Tsukuba, Tsukuba, Ibaraki 305-8577,
 Japan}}
       
\begin{document}
\begin{abstract}
We report on the status of the International Lattice Data Grid.
\end{abstract}
\maketitle

\section{Introduction}
International Lattice Data Grid (ILDG)\cite{ILDG} is an effort toward building
a data grid of lattice QCD gluon configurations  
so that people in the lattice field theory community could share them and 
fully exploit their physics content.   
It should provide an infrastructure for international research effort, 
and hopefully work as a vehicle for enhancing collaborations 
and exchange of people in our community. 

ILDG started with the proposal of the UKQCD Collaboration at 
Lattice 2002\cite{ILDG02}.  The first ILDG Workshop~\cite{ILDG-WS} 
was held in December 2002.  Discussions on the target and 
strategy of ILDG was made, and a Metadata Working Group and a 
Middleware Working Group were set up to carry out technical work.  
The second workshop followed in May 2003\cite{ILDG-WS}, and 
the ``virtual'' format of this workshop series  was established. 

The activities of ILDG in its first year was reported at 
Lattice 2003\cite{ILDG03}.  A draft version of QCDml, 
an XML for describing gluon configurations, was presented.

In this second year, two more workshops were held, in December 
2003\cite{ILDG-WS} and May 2004\cite{ILDG-WS}.
Thanks to the effort of the Metadata Working Group and people who 
participated in the discussions, v1.1 of QCDml has been 
completed.  Building up middleware also made progress.  In addition to 
technical work, the organizational aspects of ILDG was also discussed, and 
the ILDG Board was introduced at the 3rd ILDG Workshop in December 2003. 

Our report of ILDG at Lattice 2004 consists of four presentations.
The present report provides an overview of the ILDG activities since 
Lattice 2003.   
The status of the Middleware Working Group is reported by 
B\'alint Jo\'o\cite{Joo}, 
and that of the Metadata Working Group with emphasis on QCDml v1.1 
by Dirk Pleiter.  Chris Maynard 
gives a tutorial on QCDml v1.1 so that people understands it and use it 
for marking up gluon configurations.  The presentations of Pleiter and Maynard 
are combined in a single writeup\cite{Pleiter-Maynard}.

\section{Activities of Working Groups}

The Metadata Working Group consists of 
G.~Andronico (INFN), 
P.~Coddington (Adelaide),
R.~Edwards (JLAB), 
B.~Jo\'o (Edinburgh),
C.~Maynard (Edinburgh), 
D.~Pleiter (NIC/DESY), 
J.~Simone (FNAL), 
and T.~Yoshi\'e (Tsukuba, convenor).
In response to 
the presentation of QCDml draft v4.0 at Lattice 2003, important comments 
were made by the SciDAC Software Group of USA\cite{SciDAC}.  
QCDml v1.1 presented at Lattice 2004\cite{Pleiter-Maynard} incorporates these 
comments. 

While QCDml v1.1 provides a standard for the {\it description} of gluon 
configuration files, a separate standard is necessary for the {\it 
data format} of configuration files.  Discussions on the data format 
are being conducted jointly by the Metadata and Middleware Working Groups, 
and a final report is expected soon.

The Middleware Working Group consists of 
G.~Andronico (INFN), 
Y.~Chen (JLAB), 
A.~Gellrich (DESY),
J.~Hettrick (NERSC),
D.~Holmgren (FNAL),
A.~Jackson (EPCC Edinburgh), 
B.~Jo\'o (EPCC Edinburgh, co-convenor),
E.~Neilsen (FNAL),
T.~Perelmutov (FNAL),
J.~Perry (EPCC Edinburgh),
M.~Sato (Tsukuba, co-convenor),
J.~Simone (FNAL) and  
C.~Watson (JLAB, co-convenor).
The goal of this working group is to design standard middleware 
for search and retrieval of configuration files stored in ILDG.  
The architecture for the middleware is now formulated, and work is in 
progress, albeit slowly, to fill out the technical details\cite{Joo}.
   
\section{ILDG Board}

As various issues began to proliferate on technical and 
strategic aspects of ILDG, it was felt that an organizational structure 
is needed to coordinate its effort.
The ILDG Board was introduced at the 3rd ILDG Workshop to meet 
this need. Further details of the Board is as follows:\\
1. The Board consists of one member from each country to decide
policy and oversee the working groups. The initial membership is Brower
(US), Jansen (Germany), Kenway (UK), Ukawa (Japan).\\
2.  The chairperson rotates on an annual basis.  
Kenway effectively acted as chairperson till December 2003,
and Ukawa is serving for 2004.\\
3. The Board is charged with expanding the membership, seeking the
allocation of resources from their national projects, and considering
the access policy for ILDG data and guidelines for data sharing.\\
4. The chairperson is responsible for organizing the 6 monthly
workshops.

\section{Data Sharing Policy}

Up to now, only two cites offer gluon configurations 
to lattice field theory community.  One is the NERSC cite\cite{NERSC} 
set up many years ago where configurations generated by 
the MILC Collaborations are stored.  The other is the Lattice QCD Archive 
at Tsukuba \cite{LQA} opened in February 2004  
where two-flavor configurations are made available.

ILDG envisages that many more cites begins to operate in Europe and USA in 
the near future. Hence discussions and 
coordination on the data sharing policy is an important step
to make ILDG an effective tool for our community.

At the Lattice 2004 presentation of ILDG, comments were invited on this 
point.  While an undercurrent seemed to be there that an open policy is 
desirable and appreciation was expressed toward the collaborations which 
already made the configurations available, it was felt that further 
discussions were needed.

The ILDG Board discussed this issue after Lattice 2004.  It came to conclude 
that an initial data sharing policy should be moderate. 
The proposal from the ILDG Board reads as follows: \\
{\bf ILDG Data Sharing Policy}\\
In addition to the normal practice of sharing data within restricted
groups for specific joint projects, collaborations that are generating
substantial sets of gauge configurations should\\ 
1. mark up their data using the QCDml standard;\\
2. adopt a policy to make their data generally available as soon as
possible;\\
3. announce on the ILDG web pages, at the time of production, their
chosen action and parameter values, and when their configurations will
be made generally available through ILDG.

\section{Conclusions}

We are now entering a new era when new machines such as QCDOC and 
ApeNEXT begin to operate.
We hope that ILDG develops timely to exploit the potential of these and future 
machines for further progress of lattice field theory.

\end{document}